\documentclass[11pt,a4paper]{article}

\def\AmSTeX{\leavevmode\hbox{$\mathcal A\kern-.2em\lower.376ex%
        \hbox{$\mathcal M$}\kern-.2em\mathcal S$-\TeX}}
\usepackage{graphicx}
\usepackage{amsmath}        
\usepackage{amssymb}
\usepackage{mathrsfs}       
\usepackage{bm}             
\usepackage[amsmath,thmmarks,hyperref]{ntheorem}
\newif\ifpdf \pdftrue
\ifx\pdfoutput\undefined \pdffalse \fi \ifx\pdfoutput\relax
\pdffalse \fi
\ifx\texonly\undefined\let\texonly\relax\fi
\ifx\endtexonly\undefined\let\endtexonly\relax\fi \texonly
  \let\htmlonly\iffalse
  \let\endhtmlonly\fi
\endtexonly

\htmlonly
  
\endhtmlonly
\texonly
  \usepackage{makeidx}

  \usepackage{multicol}
  

\endtexonly
\title{}
\author{\thanks{}}
\date{}
\begin{document}

\title{Probing Non-leptonic Two-body Decays of $B_c$ meson}

\author{Hui-feng Fu$^a$,~~ Yue Jiang$^a$,~~ C. S. Kim$^b$\footnote{cskim@yonsei.ac.kr},~~ Guo-Li Wang$^a$\footnote{gl\_wang@hit.edu.cn}\\
{\it \small  $^a$ Department of Physics, Harbin Institute of
Technology, Harbin, 150001, China} \\
{\it \small  $^b$ Department of Physics $\&$ IPAP, Yonsei
University, Seoul 120-749, South Korea}}

\maketitle

\baselineskip=20pt
\begin{abstract}
\noindent Rates and CP asymmetries of the non-leptonic two-body
decay of $B_c$ are calculated based on the low energy effective
Hamiltonian. We concentrate on such $b$ quark decays of the
processes with $0^-$ and $1^-$ S-wave particles and/or $0^+$ and
$1^+$ P-wave particles in the final states. The  Salpeter method,
which is the relativistic instantaneous approximation of the
original Bethe-Salpeter equation, is used to derive hadron
transition matrix elements. Based on the calculation, it is found
that the best decay channels to observe CP violation  are
$B_c^-\rightarrow \eta_c +D^-(D^{*-}_0)$, which need about
$\sim10^7$ $B_c$ events in experiment. Decays to $ \eta_c
+D^{*-}(D_s^-),~~ h_c+D^-(D^{*-})(D^{*-}_0),~~ J/\Psi+D^{*-}$ are
also hopeful channels.
\end{abstract}

\section{Introduction}

The discovery of the $B_c$ meson~\cite{Abe} has provided a new
valuable window for studying the heavy quark dynamics and CP
violation. Studies on the $B_c$ decays and their CP asymmetries have
drawn much attention in accordance with the coming LHC-b experiment. Since the LHC-b is expected to
produce around $5\times10^{10}$ $B_c$ events per year~\cite{Gouz}
and to provide detailed information about the $B_c$ meson, it becomes more
and more strongly relevant to investigate $B_c$ decay and its CP
violation in detail.

There are two major reasons which make the $B_c$ meson special. The
first is that it is unique to have two heavy-flavored quarks, composed of a
charm quark (anti-quark) and a bottom anti-quark (quark). The other
heavy quark in the Standard Model, $i.e.$ the top quark, cannot form a hadron
because of its too short lifetime to be hadronized. The second reason is
that it can decay only via weak interactions, since the pure strong
and electromagnetic interacting processes conserve flavors, and the
$B_c$ meson, as the ground state of $c\bar{b}$ system, is below the
$BD$ mesons decay threshold. Due to these properties, the $B_c$  meson has a
long lifetime and rich decay channels.

The quark diagrammatic approach has established and
well developed for meson decays. In the approach, there are five diagrams
contributing to $B_c$ decays: the color-favored tree diagram, the
color-suppressed tree diagram, the time-like penguin diagram, the
annihilation diagram and the space-like penguin diagram. The direct
CP violation requires at least two diagrams with different weak
and strong phases contributing to the relevant
process. The weak phases come from CKM matrix elements within the Standard Model, and the
strong phases arise from final state interactions including penguin
effects (hard strong phases), which can be estimated perturbatively,
as well as rescattering effects (soft strong phases), which
cannot be estimated solidly now. Therefore, we only discuss penguin
effects for the generation of strong phases in this paper.

\begin{figure}[tb]
\centering
\includegraphics[width = 0.8\textwidth]{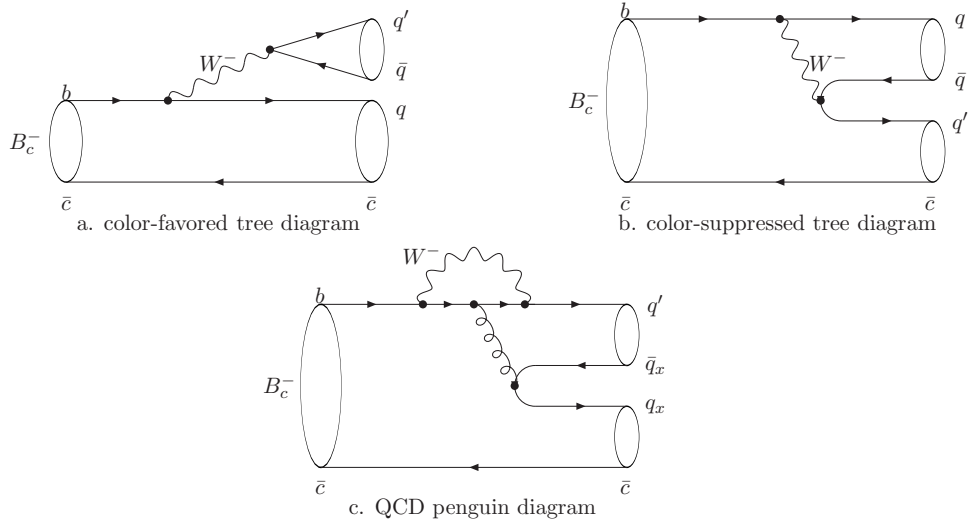}
\caption{The tree and QCD penguin diagrams of non-leptonic two-body
$B_c$ decay. $q=u$ or $c$ and $q'=s \ \mathrm{or} \ d$; $q_x$
ranges from $u,d,s$ to $c$. } \label{fig1}
\end{figure}

Non-leptonic two-body decays can play an important role for exploring
the direct CP violation. So far many works on non-leptonic $B_c$
decays and their CP violations have been
investigated~\cite{Chang}--\cite{Choi}.
But in those works CP violation of the channels with P-wave final
states has not been considered. Here we are going to concentrate on such
non-leptonic two-body decay channels that may have direct CP
asymmetries: We study the $b(\bar{b})$ quark decays with final
states involving not only pseudoscalar ($0^-$) and vector ($1^-$)
particles but also $0^+$ and $1^+$ P-wave particles.
Since the contributions from the annihilation diagram
and space-like penguin diagram are helicity suppressed, these two type
diagrams are ignored in our calculation. Furthermore, the electroweak penguin effects
are much  smaller compared to the QCD penguin effects, so only the QCD penguin
effects are considered here. Therefore, only the two tree diagrams and
the time-like QCD penguin diagrams are fully considered (see
Fig.~\ref{fig1}). The CP asymmetries arise from the interference
between the penguin diagrams and tree diagrams or/and the penguin
diagrams themselves.

In our calculation the factorization approach is assumed and the
Salpeter method is used: With the factorization approach, the
amplitude can be expressed by the products of form factors and decay
constants. The Salpeter method is used to calculate the form factors
at finite recoils. In doing so, a relativistic treatment is needed
especially for $B_c\rightarrow D/D_s+X$ processes, since the $D/D_s$
mesons are bound states composed of a heavy and a light quark and
the relativistic corrections to such particles may noticeable. It is
well known that the Bethe-Salpeter (B-S) equation is a relativistic
two-body wave equation and the Salpeter method is just the
instantaneous approximation of the B-S equation. With the Salpeter
equation and well defined wave functions, we can treat the bound
states relativistically. Therefore, in our calculations the
relativistic corrections are also systematically  covered.

The remainder of this paper is organized as follows: In section 2,
the factorization approach based on the low energy effective
Hamiltonian is introduced to evaluate the decay amplitudes. Section
3 contains a brief review on the Salpeter method and our model calculation.
Section 4 is devoted to numerical results and discussions.

\section{Nonleptonic two-body decay and its CP asymmetry of $B_c$}

In weak decay analysis, the basic starting point is the effective
weak Hamiltonian~\cite{Buras}, which in the case of $b\rightarrow u$
decay is
\begin{eqnarray}
\mathcal{H}_{\mathrm{eff}}(\Delta
B=1)=\frac{G_F}{\sqrt{2}}\left\{V_{ub}V^*_{uq'}(C_1Q_1^u+C_2Q_2^u)
-\frac{\alpha_s(m_b)}{8\pi}\left(\sum_{i=u}^{c,t}V_{ib}V^*_{iq'}I_i\right)
(-\frac{Q_3}{N_c}+Q_4-\frac{Q_5}{N_c}+Q_6)\right\},\label{eq3}
\end{eqnarray}
where $Q_1^u,Q_2^u$ are the tree operators in $b\rightarrow u$
decay, which would be replaced by $Q_1^c,Q_2^c$ in $b\rightarrow c$
decay. $Q_3,Q_4,Q_5$ and $Q_6$ are the QCD penguin operators. All these
local operators are
\begin{eqnarray}\begin{aligned}\label{eq4}
Q_1^u=(\bar{q}'_\alpha u_\beta)_{V-A}(\bar{u}_\beta
b_\alpha)_{V-A},\\
Q_2^u=(\bar{q}'_\alpha u_\alpha)_{V-A}(\bar{u}_\beta
b_\beta)_{V-A},\\
Q_1^c=(\bar{q}'_\alpha c_\beta)_{V-A}(\bar{c}_\beta
b_\alpha)_{V-A},\\
Q_2^c=(\bar{q}'_\alpha c_\alpha)_{V-A}(\bar{c}_\beta
b_\beta)_{V-A},\\
Q_3=(\bar{q}'_\alpha b_\alpha)_{V-A}\sum_{q_x}(\bar{q}_{x\beta}
q_{x\beta})_{V-A},\\
Q_4=(\bar{q}'_\alpha b_\beta)_{V-A}\sum_{q_x}(\bar{q}_{x\beta}
q_{x\alpha})_{V-A},\\
Q_5=(\bar{q}'_\alpha b_\alpha)_{V-A}\sum_{q_x}(\bar{q}_{x\beta}
q_{x\beta})_{V+A},\\
Q_6=(\bar{q}'_\alpha b_\beta)_{V-A}\sum_{q_x}(\bar{q}_{x\beta}
q_{x\alpha})_{V+A},
\end{aligned}
\end{eqnarray}
where $q'=s \ \mathrm{or} \ d$,  and the subscript $\alpha,\beta$
are color indices. $q_x$ ranges from $u,d,s$ to $c$. The operator
$(\bar{\psi}_{1\alpha}
\psi_{2\beta})_{V-A}\equiv\bar{\psi}_{1\alpha}\gamma^\mu(1-\gamma_5)
\psi_{2\beta}$, and the operators with $V+A$ represent for the right-handed currents. In
Eq.~(\ref{eq3}), $C_1,~C_2$ in front of the tree operators are
Wilson coefficients. $N_c=3$ is the number of colors and $V_{qq'}$
are the CKM matrix elements. $I_i(i=u,c,t)$ are the QCD loop
integrals~\cite{CKT,Du2}:
\begin{eqnarray}\label{eq5}
&&I_{u,c}=-4\int_0^1x(1-x)\mathrm{ln}\frac{m_{u,c}-k^2x(1-x)}{m_W^2}dx,\\
&&I_t=-\frac{1}{9}+\frac{1}{6}\int_0^1\frac{(1-x)[(2+m_t^2/m_W^2)(1-x)(2+x)+12x]}{m_t^2/m_W^2+(1-m_t^2/m_W^2)x}dx,
\end{eqnarray}
where $m_{u,c,t}$ are the current quark masses; $k$ is the momentum
of the gluon in penguin diagram, see Fig~\ref{fig1} c.
Usually one takes a certain value of $k^2$ in the range
$[\frac{1}{4} m_b^2,\frac{1}{2} m_b^2]$ or $[0,
m_b^2]$~\cite{GerHou}. As argued by the authors in Ref.~\cite{Du1},
it is not a good choice to pick up a fixed value of $k^2$ for all
decay modes. In this work, we follow the simple kinematic picture
presented in Ref.~\cite{Du1} for the value of $k^2$. One can see
from Fig.~\ref{fig1} c, as c quark being a spectator, the relation
of the momenta among the quarks and gluon
$p_b=k+p_{q'}=p_{q_x}+p_{\bar{q}_x}+p_{q'}$ are hold. Since the $q'$
quark and the $\bar{q}_x$ anti-quark form a meson, noted as $X$, the
momentum of $X$ satisfies $p_X=p_{\bar{q}_x}+p_{q'}$. With these
relations one can get
$k^2=m_b^2+m_{q'}^2-2E_bE_{q'}+2|\vec{p}_b||\vec{p}_{q'}|\cos(\phi)$,
where $\phi$ is the angle between the 3-momenta of $b$ quark
$\vec{p}_b$ and $q'$ quark $\vec{p}_{q'}$ in the rest frame of the
$X$ meson. Since the angle $\phi$ is unknown, we use the averaged
value $\bar{k}^2$ to evaluate the loop-integral functions. After
all, one get
\begin{equation}\label{eq2}
\frac{\bar{k}^2}{m_b^2}=\frac{1}{2}(1+(m_{\bar{q}_x}^2-m_{q'}^2)(1-\frac{m_{\bar{q}_x}^2}{m_b^2})/m_X^2
+(m_{q'}^2+2m_{\bar{q}_x}^2-m_X^2)/m_b^2).\end{equation}

Now we turn to evaluate the decay amplitudes in factorization
approach~\cite{Ali,Neubert} and take the
$B_c^-\rightarrow\eta_c+D^-$ channel as an example. The decay
amplitude of this process is $\langle\eta_c,D^-
|\mathcal{H}_{\mathrm{eff}}|B_c^-\rangle$. First, consider the
color-favored tree diagram (see Fig.~\ref{fig1} a), where the tree
operators $Q_1^c$ and $Q_2^c$ contribute. Using the Fierz
rearrangement
\begin{eqnarray}&(\bar{\Psi}_1\Psi_2)_{V-A}(\bar{\Psi}_3\Psi_4)_{V-A}=(\bar{\Psi}_1\Psi_4)_{V-A}(\bar{\Psi}_3\Psi_2)_{V-A},\\
&(\bar{\Psi}_{1\alpha}\Psi_{2\beta})_{V-A}(\bar{\Psi}_{3\beta}\Psi_{4\alpha})_{V-A}=
\frac{1}{N_c}(\bar{\Psi}_{1\alpha}\Psi_{2\alpha})_{V-A}(\bar{\Psi}_{3\beta}\Psi_{4\beta})_{V-A}+\mathrm{Octet},\\
&(\bar{\Psi}_1\Psi_2)_{V-A}(\bar{\Psi}_3\Psi_4)_{V+A}=-2(\bar{\Psi}_1\Psi_4)_{S+P}(\bar{\Psi}_3\Psi_2)_{S-P},
\end{eqnarray}
where the ``Octet" is the color-octet term which does not contribute
in the factorization approach. One can get the amplitude of the
color-favored tree diagram
$$\frac{G_F}{\sqrt{2}}V_{cb}V^*_{cd}a_1\langle\eta_c
|(\bar{c}b)_{V-A}|B_c^-\rangle\langle D^-
|(\bar{d}c)_{V-A}|0\rangle,$$ where $a_1\equiv\frac{C_1}{N_c}+C_2$.
The other two amplitudes (corresponding to Fig.~\ref{fig1} b and
Fig.~\ref{fig1} c) can be obtained in the same way. In the penguin
diagram, we will encounter the term
$(\bar{d}c)_{S+P}(\bar{c}b)_{S-P}$, where $(\bar{q}_1q_2)_{S\pm
P}=(\bar{q}_1(1\pm\gamma_5)q_2)$. To evaluate these terms, we use the
equation of motion, which gives
\begin{eqnarray}&\langle P'|\bar{q}_1q_2|P\rangle=\frac{P^\mu-P'^\mu}{m_1-m_2}\langle P'|\bar{q}_1\gamma_\mu
q_2|P\rangle,\\
&\langle
P'|\bar{q}_1\gamma_5q_2|P\rangle=\frac{P^\mu-P'^\mu}{m_1+m_2}\langle
P'|\bar{q}_1\gamma_\mu\gamma_5 q_2|P\rangle,
\end{eqnarray}
where $P$ and $P'$ are the momenta of initial and final states
respectively and $m_1,m_2$ are the current quark masses. Now we can
write the decay amplitude of the process
\begin{equation}\begin{aligned}
\mathcal{M}(B_c^-\rightarrow
\eta_c+D^-)=&\frac{G_F}{\sqrt{2}}\biggl\{\biggl[V_{cb}V^*_{cd}a_1-\frac{\alpha_s(m_b)}{8\pi}(V_{ub}V^*_{ud}I_{ut}+V_{cb}V^*_{cd}I_{ct})(1-\frac{1}{N_c^2})\times\\&
\times(1+\frac{2M_{D^-}^2}{(m_b-m_c)(m_d+m_c)})\biggl]\langle\eta_c
|(\bar{c}b)_{V-A}|B_c^-\rangle\langle D^-
|(\bar{d}c)_{V-A}|0\rangle\\& +V_{cb}V^*_{cd}a_2\langle D^-
|(\bar{d}b)_{V-A}|B_c^-\rangle\langle \eta_c
|(\bar{d}c)_{V-A}|0\rangle\biggl\},
\end{aligned}
\end{equation}
where $a_2\equiv\frac{C_2}{N_c}+C_1$ and $I_{ut}\equiv I_u-I_t;\
I_{ct}\equiv I_c-I_t$. The unitary condition,
$V_{ub}V^*_{uq'}+V_{cb}V^*_{cq'}+V_{tb}V^*_{tq'}=0$ with $q'=d\
\mathrm{or}\  s$, has been used to achieve the expression. The decay
width is $\Gamma=\frac{|\vec{p}|}{8\pi
M_{B_c}^2}\sum_\mathrm{pol}|\mathcal{M}|^2$, where $\vec{p}$ is the
3-momentum of one of the final state particles in the rest frame of
$B_c$. Generally the amplitude can be written as
\begin{equation}\mathcal{M}=V_{cb}V^*_{cq'}T_1+V_{ub}V^*_{uq'}T_2.
\end{equation}
The amplitude for CP conjugated process can be obtained by
conjugating the CKM matrix elements but not $T_1$ and $T_2$, $i.e.$
$\bar{\mathcal{M}}=V^*_{cb}V_{cq'}T_1+V^*_{ub}V_{uq'}T_2$.

The CP
asymmetry is defined as
\begin{equation}\mathcal{A}_{cp}=
\frac{\Gamma(B_c^+\rightarrow \bar{f})-\Gamma(B_c^-\rightarrow
f)}{\Gamma(B_c^+\rightarrow \bar{f})+\Gamma(B_c^-\rightarrow
f)}.\end{equation} Inserting the expression of the amplitude, one
can get
\begin{equation}\mathcal{A}_{cp}=\frac{\sum\biggl[2i\mathrm{Im}(T_1T_2^*)(\frac{V_{ub}V^*_{uq'}}{V_{cb}V^*_{cq'}}-(\frac{V_{ub}V^*_{uq'}}{V_{cb}V^*_{cq'}})^*)\biggl]}
{\sum\biggl[2|T_1|^2+2|\frac{V_{ub}V^*_{uq'}}{V_{cb}V^*_{cq'}}|^2|T_2|^2+2\mathrm{Re}(T_1T_2^*)(\frac{V_{ub}V^*_{uq'}}{V_{cb}V^*_{cq'}}
+(\frac{V_{ub}V^*_{uq'}}{V_{cb}V^*_{cq'}})^*)\biggl]}
..\end{equation} In the Wolfenstein parameterization of CKM matrix,
up to the $\lambda^3$ order, only $V_{ub}$ has weak phase, so we
take
$\gamma\equiv\arg(-\frac{V^*_{ub}V_{ud}}{V^*_{cb}V_{cd}})\simeq\arg(\frac{V^*_{ub}V_{us}}{V^*_{cb}V_{cs}}).$
Then the CP asymmetry drops to a simple form:
\begin{equation}\begin{aligned}
\mathcal{A}_{cp}&=\frac{\epsilon_i2\sum\mathrm{Im}(T_1T_2^*)\sin\gamma}
{\sum|T_1|^2/B_i+B_i\sum|T_2|^2+\epsilon_i2\sum\mathrm{Re}(T_1T_2^*)\cos\gamma}\\
&\equiv D_1\frac{\sin\gamma}{1+D_2\cos\gamma},\label{eq1}
\end{aligned}\end{equation}
where $\epsilon_1=+1,\epsilon_2=-1$ corresponding to
$B_1=|\frac{V_{ub}V^*_{us}}{V_{cb}V^*_{cs}}|,B_2=|\frac{V_{ub}V^*_{ud}}{V_{cb}V^*_{cd}}|$.

In our calculation, we take numerical values of CKM elements as~\cite{PDG}
\begin{eqnarray}
|V_{ud}|=0.97425,&|V_{us}|=0.2252,&|V_{ub}|=3.89\times10^{-3},\notag\\
|V_{cd}|=0.230,&|V_{cb}|=0.0406,&|V_{cs}|=0.97345.
\end{eqnarray}
For current quark masses and QCD coupling constant, we
take~\cite{CKT} $(m_u,m_d,m_s,m_c,m_b,m_t)=(0.005, 0.01,0.175,
1.35,4.8, 176)$(GeV) and $\alpha_s(m_b)=0.235$.

\section{The Salpeter method and the model calculation}

To estimate the decay rates and CP asymmetries, the
hadron matrix elements need to be calculated. In our work, we use
the Salpeter method~\cite{Salpeter2}, which is the relativistic
instantaneous approximation of the Bethe-Salpeter (B-S) equation, with well defined wave
functions to deal with the hadron matrix elements.

The B-S equation~\cite{Salpeter1} is written as
\begin{equation}\label{eq10}
(\not\!p_{1}-m_{1})\chi_{p}(q)(\not\!p_{2}+m_{2})=i\int\frac{d^{4}k}{(2\pi)^{4}}V(P,k,q)\chi_{p}(k),
\end{equation}
where $\chi_{p}(q)$ is B-S wave function of the relevant bound
state. $P$ is the four momentum of the state and $p_{1}$, $p_{2}$,
$m_{1}$, $m_{2}$ are the momenta and constituent masses of the quark
and anti-quark, respectively. From the definition
$$p_1 = \alpha_1P +
q,\  \alpha_1 \equiv \frac{m_1}{ m_1 + m_2} ,$$
$$ p_2 = \alpha_2P-q,\  \alpha_2\equiv\frac{ m_2}{ m_1 + m_2} ,$$
one can deduce the expression of relative momentum between quark and
anti-quark $q$. $V(P,k,q)$ is the interaction kernel which can be
treated as a potential after doing instantaneous approximation, $i.e.$
the kernel takes the simple form (in the rest frame)
$$V(P,k,q)\Rightarrow V(|\vec{k}-\vec{q}|).$$
For convenience, we divide the relative momentum $q$ into two parts,
$$q^\mu=q^\mu_\parallel+q^\mu_\perp,\ \ q^\mu_\parallel\equiv P\cdot q/M^2 P^\mu,\ \ q^\mu_\perp\equiv q^\mu-q^\mu_\parallel,$$
where $M$ is the mass of the meson. Correspondingly, we have two
Lorentz invariant variables:
$$q_{_P}\equiv P\cdot q/M,\ \ q_{_T}\equiv \sqrt{-q_\perp^2}.$$

With the definitions
$$ \varphi_{_P}(q_\perp^{\mu})\equiv i\int
\frac{\mathrm{d}q_{_P}}{2\pi}\chi
_{_P}(q_\parallel^{\mu},q_\perp^{\mu}),\ \ \eta(q_\perp^{\mu})\equiv
\int
\frac{\mathrm{d}k_\perp^{3}}{(2\pi)^{3}}V(k_\perp,q_\perp)\varphi_{_P}(q_\bot^{\mu}),$$
and after performing the integration over $q_P$ in Eq.
(\ref{eq10}), the B-S equation can be written as
\begin{equation}
\varphi_{_P}(q_\perp)=\frac{\Lambda_{1}^{+}(q_\perp)\eta(q_\perp)
\Lambda_{2}^{+}(q_\perp)}{M-\omega_{1}-\omega_{2}}-\frac{\Lambda_{1}^{-}(q_\perp)\eta(q_\perp)
\Lambda_{2}^{-}(q_\perp)}{M+\omega_{1}+\omega_{2}},
\end{equation}
where $\omega_{1}=\sqrt{m_{1}^{2}+q_{_T}^{2}}$,
$\omega_{2}=\sqrt{m_{2}^{2}+q_{_T}^{2}}$, and $\Lambda_{1}^{\pm},\
\Lambda_{2}^{\pm}$ are the generalized projection operators,
$$
\Lambda_{1}^{\pm}(q_\perp)\equiv
\frac{1}{2\omega_{1}}[\frac{\not\!{P}}{M}\omega_{1}\pm(m_{1}+\not\!q_\perp)],\
\ \ \  \Lambda_{2}^{\pm}(q_\perp)\equiv
\frac{1}{2\omega_{2}}[\frac{\not\!{P}}{M}\omega_{2}\mp(m_{2}+\not\!q_\perp)].
$$
Now we introduce the notations
$$\varphi_{_P}^{\pm\pm}(q_\perp)\equiv\Lambda_{1}^{\pm}(q_\perp)\frac{\not\!{P}}{M}\varphi_{_P}(q_\perp)\frac{\not\!{P}}{M}\Lambda_{2}^{\pm}(q_\perp).$$
With these notations the full Salpeter equation can be written as
\begin{eqnarray}\label{eq11}
&&(M-\omega_{1}-\omega_{2})\varphi_{_P}(q_\perp)^{++}=\Lambda_{1}^{+}(q_\perp)\eta(q_\perp)
\Lambda_{2}^{+}(q_\perp),\notag\\
&&(M+\omega_{1}+\omega_{2})\varphi_{_P}(q_\perp)^{--}=-\Lambda_{1}^{-}(q_\perp)\eta(q_\perp)
\Lambda_{2}^{-}(q_\perp),\notag\\
&&\varphi_{_P}(q_\perp)^{+-}=0,\ \ \varphi_{_P}(q_\perp)^{-+}=0.
\end{eqnarray}
In our model, the Cornell potential, which is a linear scalar
interaction plus a vector interaction, is chosen as the
instantaneous interaction kernel $V$.

In solving the equations, the constituent quark masses are taken as
$$\begin{array}{ccccc}m_u=0.305~\mathrm{GeV},&m_d=0.311~\mathrm{GeV},&m_s=0.5~\mathrm{GeV},&
m_c=1.62~\mathrm{GeV},&m_b=4.96~\mathrm{GeV}.\end{array}$$ The form
of wave functions with certain quantum numbers
$J^{P(C)}=0^{-(+)},1^{-(-)},0^{+(+)},1^{+(+)}$ and $1^{+(-)}$\footnote{
$J^P$ for general particles, $J^{PC}$ for quarkonium $i.e.$ the equal mass
system. The wave functions satisfy the correct $C$-parity
spontaneously when the masses of quark and anti-quark are equal.}
are written as
\begin{eqnarray}\label{eq12}
\varphi_{0^{-(+)}}(q_\perp)&=&M\left[\frac{\not\!P}{M}a_1(q_\perp)+{a}_{2}(q_\perp)
+\frac{\not\!q_\perp}{M}{a}_{3}(q_\perp)+\frac{\not\!{P}\not\!q_\perp}{M^2}{a}_{4}(q_\perp)\right]{\gamma}_5,
\notag\\
{\varphi}_{1^{-(-)}}({q_\perp})&=&(q_{\perp}\cdot{\epsilon}^{\lambda}_{\perp})\left[b_1({q_\perp})+\frac{\not\!{P}}{M}b_2({q_\perp})
+\frac{\not\!q_{\bot}}{M}b_3({q_\perp})+\frac{\not\!{P}\not\!{q_{\bot}}}{M^2}b_4({q_\perp})\right]+M\not\!{\epsilon}^{\lambda}_{\bot}b_5({q_\perp})\notag\\
&&+\not\!{\epsilon}^{\lambda}_{\bot}\not\!{P}b_6(\vec{6})
+(\not\!{q_{\bot}}\not\!{\epsilon}^{\lambda}_{\bot}-q_{\bot}\cdot{\epsilon}^{\lambda}_{\bot})b_7({q_\perp})
+\frac{1}{M}(\not\!{P}\not\!{\epsilon}^{\lambda}_{\bot}\not\!{q_{\bot}}-\not\!{P}q_{\bot}\cdot{\epsilon}^{\lambda}_{\bot})b_8({q_\perp}),\notag\\
{\varphi}_{0^{+(+)}}({q_\perp})&=&{f}_{1}({q_\perp})\not\!{q_{\bot}}+{f}_{2}({q_\perp})\frac{\not\!{P}\not\!{q_{\bot}}}{M}
+{f}_{3}({q_\perp})M+{f}_{4}({q_\perp})\not\!{P},\notag\\
{\varphi}_{1^{+(+)}}({q_\perp})&=&i{\varepsilon}_{\mu\nu\alpha\beta}P^{\nu}q^{\alpha}_{\bot}{\epsilon}^{\lambda\beta}_\perp
\biggl[g_1({q_\perp})M{\gamma}^{\mu}+g_2({q_\perp})\not\!{P}{\gamma}^{\mu}+g_3({q_\perp})\not\!{q_{\bot}}{\gamma}^{\mu}\notag\\
&&+ig_4({q_\perp}){\varepsilon}^{\mu\rho\sigma\delta}P_{\sigma}q_{\bot\rho}{\gamma}_{\delta}{\gamma}_5/M\biggl]/M^2,\notag\\
{\varphi}_{1^{+(-)}}({q_\perp})&=&q_{\bot}\cdot\epsilon^\lambda_\perp\left[h_1({q_\perp})+h_2({q_\perp})\frac{\not\!P}{M}
+h_3({q_\perp})\not\!{q_{\bot}}+h_4({q_\perp})\frac{\not\!P\not\!{q_{\bot}}}{M^2}\right]{\gamma}_5,
\end{eqnarray}
where $a_i(q_\perp),b_i(q_\perp),f_i(q_\perp),g_i(q_\perp)$ and
$h_i(q_\perp)$ are wave functions to $q_\perp^2$; $M$ is the mass of
corresponding bound state; $\epsilon^\lambda_\perp$ is the
polarization vector for $J^P=1^{\pm}$ state. With these wave
functions, we solve the Salpeter equation (Eq.~(\ref{eq11})) and get
$$\begin{array}{cccc}
M_{\bar{D}^0}=1.865,&M_{D^-}=1.869,&M_{D_s^-}=1.968,&M_{\eta_c}=2...980,\\
M_{\bar{D}^{*0}}=2.006,&M_{D^{*-}}=2.011,&M_{D_s^{*-}}=2.112,&M_{J/\Psi}=3.097,\\
M_{\bar{D}_0^{*0}}=2.317,&M_{D_0^{*-}}=2.323,&M_{D_{s0}^{*-}}=2.318,&M_{\chi_{c0}}=3.415,\\
M_{\chi_{c1}}=3.510,&M_{h_c}=3.526,&M_{D_{s1}^-(2460)}=2.459,&M_{D_{s1}^-(2536)}=2.535,
\end{array}$$
and $M_{B_c}=6.276$ in unit of GeV. In our method, the wave
functions are constructed for certain $J^{PC}$ quantum state, such
as $\chi_{c1}$, which is a $J^{PC}=1^{++}$ state and also a
$^{2s+1}L_J={^3P_1}$ state and $h_c$ which is a $1^{+-}$ or $^1P_1$
state. This is the case for quarkonium. For the particles composed
of a couple of quark and anti-quark with different masses, the two
states are just ${^3P_1}$ and ${^1P_1}$ states and both are
$J^P=1^+$ states (such states don't have C-parity), so the mixture
between the ${^3P_1}$ and ${^1P_1}$ states may happen. If one puts
the quark masses equal, the two states are spontaneously deduced to
$1^{++}$ and $1^{+-}$ states respectively. The particles
$D_{s1}^-(2460)$ and $D_{s1}^-(2536)$ are considered to be mixed of
${^3P_1}$ and ${^1P_1}$ states. In this work we take the mixing
relation as
$$|P_1^{1/2}\rangle=-\frac{1}{\sqrt{3}}|^1P_1\rangle+\sqrt{\frac{2}{3}}|^3P_1\rangle,\hspace{15mm}
|P_1^{3/2}\rangle=\sqrt{\frac{2}{3}}|^1P_1\rangle+\frac{1}{\sqrt{3}}|^3P_1\rangle,$$
where $|P_1^{1/2}\rangle$ corresponds to the $D_{s1}^-(2460)$ and
$|P_1^{3/2}\rangle$ corresponds to the $D_{s1}^-(2536)$. Interested
reader can find details about the Salpeter method and our model in
Ref.~\cite{Wang}.

With the wave functions of bound states, we can calculate hadron
matrix elements, such as $\langle\eta_c
|(\bar{c}b)_{V-A}|B_c^-\rangle\langle D^-
|(\bar{d}c)_{V-A}|0\rangle$. According to Mandelstam
formalism~\cite{Mandelstam}, at the leading order, the transition
matrix element can be written as~\cite{Wang2}
\begin{equation}
\langle\eta_c |(\bar{c}\Gamma^\mu
b)|B_c^-\rangle=\int\frac{d^3q_\perp}{(2\pi)^3}\mathrm{Tr}\Big[\bar{\varphi}_{\eta_c}^{++}(q_\perp+\alpha_2'P'_\perp)
\Gamma^\mu\varphi_{B_c^-}^{++}(q_\perp)\frac{\not\!P}{M}\Big],
\end{equation}
where $\Gamma^\mu=\gamma^\mu(1-\gamma_5)$; $P$ and $M$ is the
momentum and mass of initial state, $i.e.$ the $B_c$ meson; $P'$ is
the momentum of $\eta_c$ and $P'_\perp=\frac{P'\cdot P}{M^2} P$;
$\alpha_2'=\frac{m_c}{m_c+m_c}$; and
$\bar{\varphi}_{\eta_c}^{++}=\gamma_0\varphi_{\eta_c}^{++}\gamma_0$.
The $\langle D^- |(\bar{d}c)_{V-A}|0\rangle$ is just a decay
constant. For $J^P=0^\pm$ and $1^\pm$ particles, we define decay
constants $f_{0^\pm}$ and $f_{1^\pm}$ as
\begin{eqnarray}\langle
P(0^\pm)
|(\bar{q}_1q_2)_{V-A}|0\rangle\equiv if_{0^\pm} P^\mu, \\
\langle P(1^\pm) |(\bar{q}_1q_2)_{V-A}|0\rangle\equiv
if_{1^\pm}M\epsilon^\mu. \end{eqnarray} Accordingly the transition
matrix elements can be expressed with form factors:
\begin{eqnarray}&\langle
P'(0^\pm)
|(\bar{q}_1q_2)_{V-A}|P(B_c^-)\rangle\equiv f_+(P+P')^\mu+f_-(P-P')^\mu, \\
&\langle P'(1^\pm) |(\bar{q}_1q_2)_{V-A}|P(B_c^-)\rangle\equiv
f_1\frac{\epsilon\cdot P}{M} P^\mu +f_2\frac{\epsilon\cdot P}{M}
P'^\mu +f_3\epsilon^\mu + if_4\varepsilon^{\mu\epsilon P P'},
\end{eqnarray}
where $f_\pm$ and $f_i\ (i=1,2,3,4)$ are form factors. After all the
hadron matrix can be expressed in the products of decay constants
and form factors.

\section{Numerical results and discussions}

We now use the method previously illustrated to estimate the non-leptonic two-body decay widths of
$B_c$  meson and their CP asymmetries. In our
calculation the decay constants are taken from experimental values or
Lattice QCD results, if available. Otherwise, we use the values shown in   Table \ref{tab1}.

\begin{table} \caption{Decay constants used in our calculation in unit of MeV.}
\begin{tabular}{|c|c|c|c|c|c|c|c|c|} \hline\hline$f_{\pi^-}$&$f_{k^-}$&$f_{D^+}$&$f_{D_s^+}$&$f_{\rho}$&$f_{k^*}$
&$f_{\phi}$&$f_{D^*}$&$f_{D_s^*}$\\\hline
130 \cite{PDG}&156 \cite{PDG}&207 \cite{PDG}&258 \cite{PDG}&205 \cite{Ball}&217 \cite{Ball}&231 \cite{Ball}&245 \cite{Bec}&272 \cite{Bec}\\
\hline\hline$f_{J/\Psi}$&$f_{\eta_c}$&$f_{D^*_0}$&$f_{D_{s0}^*}$&$f_{D_{s1}(2360)}$&$f_{D_{s1}(2536)}$
&$f_{\chi_{c0}}$&$f_{\chi_{c1}}$&$f_{h_c}$\\\hline
409\cite{Bram}&420&137&109&227&77.3&0&239&0\\\hline
\end{tabular}\label{tab1}
\end{table}

The decay widths of $B_c^-\rightarrow \eta_c(J/\Psi) + X^-$ for
general values of the Wilson coefficients $a_1$ and $a_2$ are
tabulated in Table \ref{tab2} compared with the results from other
models. We can see that the results from different models are
roughly comparable. In our calculation the penguin contributions are
shown explicitly keeping the weak phase free. Then it is easy to say
that the CP violation arises from the interference between the term
with $e^{-i\gamma}$, which is the penguin contribution, and the terms
without it, which are dominated by the tree contribution.

\begin{table} \scriptsize\caption{The decay widths of $B_c^-\rightarrow \eta_c(J/\Psi) + X^-$
in the unit of $10^{-15}$ GeV for general values of the Wilson coefficients $a_1$, $a_2$ and weak
phase $\gamma$.}
\begin{tabular}{|c|c|c|c|c|}
\hline Final States& Ours
&Hern$\acute{\mathrm{a}}$ndez $et.~al.$ \cite{Hern}&Hady $et.~al.$ \cite{Hady}&Ivanov $et.~al.$ \cite{Ivan2}
\\\hline $\eta_c+D^-$
&$|0.438a_1+0.290a_2-(0.0487-0.0174
i)$&$(0.438a_1+0.236a_2)^2$&$(0.485a_1+0.528a_2)^2$&$(0.562a_1+0.582a_2)^2$
\\&$+(0.0185-0.00770
i)\textrm{e}^{-i\gamma}|^2$&&&\\\hline $\eta_c+D^{*-}$
&$|0.431a_1+0.329a_2-(0.0196-0.00681
i)$&$(0.390a_1+0.136a_2)^2$&$(0.466a_1+0.452a_2)^2$&$(0.511a_1+0.310a_2)^2$
\\&$+(0.00673-0.00305
i)\textrm{e}^{-i\gamma}|^2$&&&\\\hline $\eta_c+D_0^{*-}$
&$|0.293a_1+0.289a_2-(0.0458-0.0149 i)$&&&
\\&$+(0.0156-0.00691
i)\textrm{e}^{-i\gamma}|^2$&&&\\\hline $J/\psi+D^{-}$
&$|0.371a_1+0.258a_2-(0.00273-0.000978
i)$&$(0.328a_1+0.156a_2)^2$&$(0.372a_1+0.338a_2)^2$&$(0.462a_1+0.277a_2)^2$
\\&$+(0.00113-0.000432
i)\textrm{e}^{-i\gamma}|^2$&&&\\\hline $J/\psi+D_0^{*-}$
&$|0.212a_1+0.273a_2+(0.00307-0.00100 i)$&&&
\\&$-(0.00105-0.000463
i)\textrm{e}^{-i\gamma}|^2$&&&\\
\hline\hline $\eta_c+D_s^-$ &$|2.32a_1+1.81a_2-(0.259-0.0909
i)$&$(2.54a_1+1.93a_2)^2$&$(2.16a_1+2.57a_2)^2$&$(2.73a_1+2.82a_2)^2$
\\&$-(0.00471-0.00221
i)\textrm{e}^{-i\gamma}|^2$&&&\\\hline $\eta_c+D_s^{*-}$
&$|1.98a_1+1.82a_2-(0.0909-0.0309
i)$&$(1.84a_1+1.17a_2)^2$&$(2.03a_1+2.16a_2)^2$&$(2.29a_1+1.51a_2)^2$
\\&$-(0.00177-0.000764
i)\textrm{e}^{-i\gamma}|^2$&&&\\\hline $\eta_c+D_{s0}^{*-}$
&$|0.987a_1+1.34a_2-(0.169-0.0550 i)$&&&
\\&$-(0.00320-0.00139
i)\textrm{e}^{-i\gamma}|^2$&&&\\\hline $\eta_c+D_{s1}^{-}(2460)$
&$|1.45a_1+1.70a_2-(0.0691-0.0218 i)$&&&
\\&$-(0.00129-0.000561
i)\textrm{e}^{-i\gamma}|^2$&&&\\\hline $\eta_c+D_{s1}^{-}(2536)$
&$|0.475a_1-1.59a_2-(0.0227-0.00704 i)$&&&
\\&$-(0.000450-0.000183
i)\textrm{e}^{-i\gamma}|^2$&&&\\\hline $J/\psi+D_s^-$
&$|1.92a_1+1.52a_2-(0.0151-0.00528
i)$&$(1.85a_1+1.23a_2)^2$&$(1.62a_1+1.72a_2)^2$&$(2.19a_1+1.32a_2)^2$
\\&$-(0.000274-0.000129
i)\textrm{e}^{-i\gamma}|^2$&&&\\\hline $J/\psi+D_{s0}^{*-}$
&$|0.714a_1+1.29a_2+(0.0163-0.00531 i)$&&&
\\&$+(0.000309-0.000134
i)\textrm{e}^{-i\gamma}|^2$&&&\\\hline
\end{tabular} \label{tab2}
\end{table}

The decay widths of $B_c^-\rightarrow \chi_{c0}/\chi_{c1}/h_c + X^-$
for general values of the Wilson coefficients $a_1$, $a_2$ and weak
phase $\gamma$ are shown in Table \ref{tab3}. For $B_c^-\rightarrow
\chi_{c0}/h_c+X^-$ decay, the contribution from the color-suppressed
diagram is vanished due to the zero decay constants of $\chi_{c0}$
and $h_c$, so the $a_1$ term dominates the decay width. But this is
not true for the decay $B_c^-\rightarrow \chi_{c1}+X^-$. We can see
from the table that the numerical factors in front of $a_2$ are
about several times as the factors in front of $a_1$, and because
the Wilson coefficients are usually taken as $a_1=1.14,\
a_2=-0.20$~\cite{Bram}, the two terms are in the same order and may
cancel each other a lot. It means that in this case the decay widths
are largely suppressed and may cover effective contributions from
the penguin diagrams.

\begin{table}[ph]
\caption{The decay widths of $B_c^-\rightarrow
\chi_{c0}/\chi_{c1}/h_c + X^-$ in the unit of
$10^{-15}$ GeV for general values of the Wilson
coefficients $a_1$, $a_2$ and weak phase $\gamma$. } {\begin{tabular}{|c|c|} \hline Final States& Ours
\\\hline $\chi_{c0}+D^-$
&$|0.193a_1-(0.00142-0.000508 i)+(0.000587-0.000224
i)\textrm{e}^{-i\gamma}|^2$
\\\hline $\chi_{c0}+D^{*-}$
&$|0.224a_1-(0.0101-0.00353 i)+(0.00349-0.00158
i)\textrm{e}^{-i\gamma}|^2$
\\\hline $\chi_{c0}+D_0^{*-}$
&$|0.109a_1+(0.00158-0.000514 i)-(0.000538-0.000238
i)\textrm{e}^{-i\gamma}|^2$
\\\hline $h_c+D^-$
&$|0.292a_1-(0.0325-0.0116 i)+(0.0135-0.00514
i)\textrm{e}^{-i\gamma}|^2$
\\\hline $h_c+D^{*-}$
&$|0.290a_1-(0.0131-0.00458 i)+(0.00453-0.00205
i)\textrm{e}^{-i\gamma}|^2$
\\\hline $h_c+D_0^{*-}$
&$|0.131a_1-(0.0205-0.00668 i)+(0.00699-0.00309
i)\textrm{e}^{-i\gamma}|^2$
\\\hline $\chi_{c1}+D^-$
&$|0.0465a_1+0.173a_2-(0.00516-0.00185 i)+(0.00214-0.000817
i)\textrm{e}^{-i\gamma}|^2$
\\\hline $\chi_{c1}+D_0^{*-}$
&$|0.0217a_1+0.164a_2-(0.00339-0.00110 i)+(0.00116-0.000511
i)\textrm{e}^{-i\gamma}|^2$
\\\hline\hline $\chi_{c0}+D_s^-$
&$|0.991a_1-(0.00779-0.00273 i)-(0.000142-0.0000665
i)\textrm{e}^{-i\gamma}|^2$
\\\hline $\chi_{c0}+D_s^{*-}$
&$|1.00a_1-(0.0460-0.0157 i)-(0.000898-0.000387
i)\textrm{e}^{-i\gamma}|^2$
\\\hline $\chi_{c0}+D_{s0}^{*-}$
&$|0.367a_1+(0.00837-0.00273 i)+(0.000159-0.0000689
i)\textrm{e}^{-i\gamma}|^2$
\\\hline $\chi_{c0}+D_{s1}^-(2460)$
&$|0.631a_1-(0.0300-0.00946 i)-(0.000559-0.000243
i)\textrm{e}^{-i\gamma}|^2$
\\\hline $\chi_{c0}+D_{s1}^-(2536)$
&$|0.192a_1-(0.00917-0.00284 i)-(0.000181-0.0000739
i)\textrm{e}^{-i\gamma}|^2$
\\\hline $h_c+D_s^-$
&$|1.46a_1-(0.163-0.0572 i)-(0.00297-0.00139
i)\textrm{e}^{-i\gamma}|^2$
\\\hline $h_c+D_s^{*-}$
&$|1.26a_1-(0.0580-0.0197 i)-(0.00113-0.000487
i)\textrm{e}^{-i\gamma}|^2$
\\\hline $h_c+D_{s0}^{*-}$
&$|0.445a_1-(0.0761-0.0248 i)-(0.00144-0.000627
i)\textrm{e}^{-i\gamma}|^2$
\\\hline $h_c+D_{s1}^-(2460)$
&$|0.716a_1-(0.0340-0.0107 i)-(0.000635-0.000276
i)\textrm{e}^{-i\gamma}|^2$
\\\hline $h_c+D_{s1}^-(2536)$
&$|0.214a_1-(0.0103-0.00318 i)-(0.000203-0.0000827
i)\textrm{e}^{-i\gamma}|^2$
\\\hline $\chi_{c1}+D_s^-$
&$|0.232a_1+0.930a_2-(0.0259-0.00909 i)-(0.000471-0.000221
i)\textrm{e}^{-i\gamma}|^2$
\\\hline $\chi_{c1}+D_{s0}^{*-}$
&$|0.0740a_1+0.716a_2-(0.0127-0.00412 i)-(0.000240-0.000104
i)\textrm{e}^{-i\gamma}|^2$
\\\hline
\end{tabular} \label{tab3}}
\end{table}

The decay width of $B_c^-$ decaying into a heavy meson
($\bar{D},D^-,\dots$) and a light meson ($\pi,K,\dots$) are shown in
Table \ref{tab4}. In these decay channels, either the color-favored
tree diagram or the color-suppressed tree diagram contributes. It can be
seen from the table that the penguin diagram contribution is in the
leading order as the tree contribution in decays $B_c^-\rightarrow
\bar{D}^{(*)0}_{(0)}+K^{(*)-}$. For the decays $B_c^-\rightarrow
D^{*-}_{(0)}+\pi^0/\rho^0$, noticing $a_2\simeq -0.2$, one can find
that the penguin effects are as large as the tree diagram
contributions. So one can expect these channels have sufficiently
large CP asymmetries.

\begin{table}[tb]
\caption{The decay widths of $B_c^-\rightarrow D+$light meson in the unit of
$10^{-15}$ GeV for
general values of the Wilson coefficients $a_1$, $a_2$ and weak
phase $\gamma$. }
{\begin{tabular}{|c|c|c|} \hline Final States& Ours &Choi $et.~al.$ \cite{Choi}
\\\hline $\bar{D}^0+K^-$
&$|(0.00425a_1-(0.000286-0.000116
i))\textrm{e}^{-i\gamma}-(0.0156-0.00367 i)|^2$&$(0.00625a_1)^2$
\\\hline $\bar{D}^0+K^{*-}$
&$|(0.00621a_1-(0.000267-0.000108
i))\textrm{e}^{-i\gamma}-(0.0143-0.00367 i)|^2$&$(0.00866a_1)^2$
\\\hline $\bar{D}^{*0}+K^-$
&$|(0.00614a_1-(0.000115-0.0000466
i))\textrm{e}^{-i\gamma}-(0.00629-0.00148 i)|^2$&
\\\hline $\bar{D}^{*0}+K^{*-}$
&$|(0.00942a_1-(0.000404-0.000164
i))\textrm{e}^{-i\gamma}-(0.0217-0.00556 i)|^2$&
\\\hline $\bar{D}_0^{*0}+K^-$
&$|(0.00401a_1-(0.0000753-0.0000304
i))\textrm{e}^{-i\gamma}-(0.00411-0.000966 i)|^2$&
\\\hline $\bar{D}_0^{*0}+K^{*-}$
&$|(0.00594a_1-(0.000255-0.000103
i))\textrm{e}^{-i\gamma}-(0.0137-0.00351 i)|^2$&
\\\hline $\bar{D}^0+\pi^-$
&$|(0.0150a_1-(0.000983-0.000404
i))\textrm{e}^{-i\gamma}+(0.00284-0.000794 i)|^2$&$(0.0217a_1)^2$
\\\hline $\bar{D}^0+\rho^-$
&$|(0.0249a_1-(0.00108-0.000434
i))\textrm{e}^{-i\gamma}+(0.00308-0.000838 i)|^2$&$(0.0374a_1)^2$
\\\hline $\bar{D}^{*0}+\pi^-$
&$|(0.0219a_1-(0.000422-0.000174
i))\textrm{e}^{-i\gamma}+(0.00122-0.000341 i)|^2$&
\\\hline $\bar{D}^{*0}+\rho^-$
&$|(0.0374a_1-(0.00162-0.000652
i))\textrm{e}^{-i\gamma}+(0.00462-0.00126 i)|^2$&
\\\hline $\bar{D}_0^{*0}+\pi^-$
&$|(0.0141a_1-(0.000272-0.000112
i))\textrm{e}^{-i\gamma}+(0.000787-0.000220 i)|^2$&
\\\hline $\bar{D}_0^{*0}+\rho^-$
&$|(0.0238a_1-(0.00103-0.000415
i))\textrm{e}^{-i\gamma}+(0.00294-0.000801 i)|^2$&
\\\hline\hline $D^-+\pi^0$
&$|(0.0108a_2+(0.000649-0.000260
i))\textrm{e}^{-i\gamma}-(0.00183-0.000511 i)|^2$&$(0.0155a_2)^2$
\\\hline $D^-+\rho^0$
&$|(0.0181a_2+(0.000764-0.000315
i))\textrm{e}^{-i\gamma}-(0.00223-0.000608 i)|^2$&$(0.0265a_2)^2$
\\\hline $D^{*-}+\pi^0$
&$|(0.0157a_2+(0.000424-0.000170
i))\textrm{e}^{-i\gamma}-(0.00119-0.000333 i)|^2$&
\\\hline $D^{*-}+\rho^0$
&$|(0.0270a_2+(0.00114-0.000470
i))\textrm{e}^{-i\gamma}-(0.00333-0.000908 i)|^2$&
\\\hline $D_0^{*-}+\pi^0$
&$|(0.0102a_2+(0.000275-0.000110
i))\textrm{e}^{-i\gamma}-(0.000774-0.000216 i)|^2$&
\\\hline $D_0^{*-}+\rho^0$
&$|(0.0171a_2+(0.000723-0.000297
i))\textrm{e}^{-i\gamma}-(0.00211-0.000575 i)|^2$&
\\\hline
\end{tabular} \label{tab4}}
\end{table}

To estimate numerical values of decay rates and CP asymmetries,
we now take $a_1=1.14$ and $a_2=-0.2$~\cite{Choi,Bram}. For the weak phase,
we use the relation $\gamma=\arg(\bar{\rho}+i\bar{\eta})$ and take
the value $\bar{\rho}=0.132, \  \bar{\eta}=0.341$~\cite{PDG}, which
give $\gamma=1.20~(68.8^\circ)$. The lifetime
$\tau_{B_c}=0.46$~\cite{lifetime} is used to calculate the decay branching
ratio. Taking these values we calculate branching ratios and
CP asymmetries of non-leptonic two-body $B_c$ decay which are shown
in Table \ref{tab5} and Table \ref{tab6}. In the tables, $D_1$ and
$D_2$ are defined in Eq. (\ref{eq1}); $\epsilon_f N$ in the last column
are the numbers of $B_c^{\pm}$ events needed for testing CP
violation. For three standard deviation (3$\sigma$) signature
$\epsilon_f N\sim\frac{9}{Br \mathcal{A}^2_{cp}}$, where
$\epsilon_f$ is the detecting efficiency of the final state.

\begin{table}[ph]
\caption{The CP asymmetries and branching ratios of $B_c$
non-leptonic decays [\uppercase\expandafter{\romannumeral1}]. $D_1$
and $D_2$ are defined as $\mathcal{A}_{CP}\equiv
D_1\frac{\sin\gamma}{1+D_2\cos\gamma}$, where $\gamma$ is the weak
phase (see Eq. (\ref{eq1})); $\epsilon_f N$ are the numbers of
$B_c^{\pm}$ events needed for testing CP violation at three standard
deviation (3$\sigma$) level, which are decided by $\epsilon_f
N\sim\frac{9}{Br \mathcal{A}^2_{cp}}$, where $\epsilon_f$ is the
detecting efficiency of the final state. The following values are
taken in calculation: $a_1=1.14$, $a_2=-0.2$ and weak phase
$\gamma=68.8^{\circ}$. For channels with $\eta_c,J/\Psi$ and
$\chi_{c1}$, color-favored tree, color-suppressed tree and penguin
diagrams contribute; for channels with $\chi_{c0}$ and $h_c$ only
color-favored tree and penguin diagrams contribute due to the zero
decay constants of the $\chi_{c0}$ and $h_c$. }
{\begin{tabular}{|c|c|c|c|c|c|c|c|} \hline No.&Final
States&$D_1$&$D_2$&$\mathcal{A}_{cp}$&$B_r(B_c^+\rightarrow
\bar{f})$&$B_r(B_c^-\rightarrow f)$&$\epsilon_f N$
\\\hline 1&$\eta_c+D^-$
&0.0432&0.0920&0.0390&$1.16\times10^{-4}$&$1.07\times10^{-4}$&$5.30\times10^{7}$
\\\hline 2&$\eta_c+D_0^{*-}$
&0.0680&0.130&0.0605&$4.18\times10^{-5}$&$3.70\times10^{-5}$&$6.23\times10^{7}$
\\\hline 3&$\eta_c+D^{*-}$
&0.0155&0.0329&0.0143&$1.19\times10^{-4}$&$1.15\times10^{-4}$&$3.76\times10^{8}$
\\\hline 4&$\chi_{c0}+D^-$
&0.00207&0.00538&0.00192&$3.34\times10^{-5}$&$3.33\times10^{-5}$&$7.30\times10^{10}$
\\\hline 5&$\chi_{c0}+D_0^{*-}$
&-0.00375&-0.00859&-0.00351&$1.09\times10^{-5}$&$1.10\times10^{-5}$&$6.66\times10^{10}$
\\\hline 6&$\chi_{c0}+D^{*-}$
&0.0133&0.0283&0.0123&$4.29\times10^{-5}$&$4.18\times10^{-5}$&$1.41\times10^{9}$
\\\hline 7&$h_c+D^-$
&0.0375&0.0879&0.0339&$6.77\times10^{-5}$&$6.33\times10^{-5}$&$1.20\times10^{8}$
\\\hline 8&$h_c+D_0^{*-}$
&0.0532&0.105&0.0477&$1.27\times10^{-5}$&$1.16\times10^{-5}$&$3.25\times10^{8}$
\\\hline 9&$h_c+D^{*-}$
&0.0133&0.0283&0.0123&$7.21\times10^{-5}$&$7.03\times10^{-5}$&$8.38\times10^{8}$
\\\hline 10&$\chi_{c1}+D^-$
&0.161&0.292&0.135&$1.62\times10^{-7}$&$1.23\times10^{-7}$&$3.45\times10^{9}$
\\\hline 11&$\chi_{c1}+D_0^{*-}$
&-0.0681&-0.205&-0.0686&$8.14\times10^{-8}$&$9.33\times10^{-8}$&$2.19\times10^{10}$
\\\hline 12&$\chi_{c1}+D^{*-}$
&0.0178&0.0373&0.0164&$2.62\times10^{-5}$&$2.54\times10^{-5}$&$1.30\times10^{9}$
\\\hline 13&$J/\psi+D^{-}$
&0.00236&0.00613&0.00219&$9.55\times10^{-5}$&$9.51\times10^{-5}$&$1.96\times10^{10}$
\\\hline 14&$J/\psi+D_0^{*-}$
&-0.00482&-0.0111&-0.00451&$2.50\times10^{-5}$&$2.52\times10^{-5}$&$1.76\times10^{10}$
\\\hline 15&$J/\psi+D^{*-}$
&0.0156&0.0329&0.0143&$3.23\times10^{-4}$&$3.14\times10^{-4}$&$1.37\times10^{8}$
\\\hline\hline 16&$\eta_c+D_s^-$
&-0.00239&-0.00455&-0.00223&$2.860\times10^{-3}$&$2.873\times10^{-3}$&$6.30\times10^{8}$
\\\hline 17&$\eta_c+D_{s0}^{*-}$
&-0.00474&-0.00891&-0.00443&$3.322\times10^{-4}$&$3.352\times10^{-4}$&$1.37\times10^{9}$
\\\hline 18&$\eta_c+D_s^{*-}$
&-0.000881&-0.00195&-0.000821&$2.273\times10^{-3}$&$2.276\times10^{-3}$&$5.87\times10^{9}$
\\\hline 19&$\eta_c+D_{s1}^-(2460)$
&-0.000933&-0.00205&-0.000870&$1.091\times10^{-3}$&$1.093\times10^{-3}$&$1.09\times10^{10}$
\\\hline 20&$\eta_c+D_{s1}^-(2536)$
&-0.000447&-0.00107&-0.000416&$4.900\times10^{-4}$&$4.904\times10^{-4}$&$1.06\times10^{11}$
\\\hline 21&$\chi_{c0}+D_s^-$
&-0.000119&-0.000252&-0.000111&$8.802\times10^{-4}$&$8.804\times10^{-4}$&$8.28\times10^{11}$
\\\hline 22&$\chi_{c0}+D_{s0}^{*-}$
&0.000318&0.000746&0.000297&$1.273\times10^{-4}$&$1.272\times10^{-4}$&$8.04\times10^{11}$
\\\hline 23&$\chi_{c0}+D_s^{*-}$
&-0.000728&-0.00163&-0.000679&$8.412\times10^{-4}$&$8.423\times10^{-4}$&$2.32\times10^{10}$
\\\hline 24&$\chi_{c0}+D_{s1}^-(2460)$
&-0.000728&-0.00161&-0.000679&$3.320\times10^{-4}$&$3.325\times10^{-4}$&$5.88\times10^{10}$
\\\hline 25&$\chi_{c0}+D_{s1}^-(2536)$
&-0.000730&-0.00172&-0.000680&$3.057\times10^{-5}$&$3.062\times10^{-5}$&$6.35\times10^{11}$
\\\hline 26&$h_c+D_s^-$
&-0.00200&-0.00387&-0.00187&$1.576\times10^{-3}$&$1.581\times10^{-3}$&$1.63\times10^{9}$
\\\hline 27&$h_c+D_{s0}^{*-}$
&-0.00328&-0.00651&-0.00306&$1.298\times10^{-4}$&$1.306\times10^{-4}$&$7.36\times10^{9}$
\\\hline 28&$h_c+D_s^{*-}$
&-0.000728&-0.00163&-0.000679&$1.336\times10^{-3}$&$1.338\times10^{-3}$&$1.46\times10^{10}$
\\\hline 29&$h_c+D_{s1}^-(2460)$
&-0.000728&-0.00161&-0.000679&$4.278\times10^{-4}$&$4.284\times10^{-4}$&$4.56\times10^{10}$
\\\hline 30&$h_c+D_{s1}^-(2536)$
&-0.000730&-0.00172&-0.000680&$3.827\times10^{-5}$&$3.832\times10^{-5}$&$5.08\times10^{11}$
\\\hline 31&$\chi_{c1}+D_s^-$
&-0.0112&-0.0160&-0.0105&$1.964\times10^{-6}$&$2.005\times10^{-6}$&$4.14\times10^{10}$
\\\hline 32&$\chi_{c1}+D_{s0}^{*-}$
&0.00252&0.00686&0.00234&$3.609\times10^{-6}$&$3.592\times10^{-6}$&$4.56\times10^{11}$
\\\hline 33&$\chi_{c1}+D_s^{*-}$
&-0.000995&-0.00218&-0.000928&$4.919\times10^{-4}$&$4.928\times10^{-4}$&$2.12\times10^{10}$
\\\hline 34&$\chi_{c1}+D_{s1}^-(2460)$
&-0.00108&-0.00235&-0.00101&$1.761\times10^{-4}$&$1.764\times10^{-4}$&$5.01\times10^{10}$
\\\hline 35&$\chi_{c1}+D_{s1}^-(2536)$
&-0.000633&-0.00150&-0.000591&$4.259\times10^{-5}$&$4.264\times10^{-5}$&$6.05\times10^{11}$
\\\hline 36&$J/\psi+D_s^-$
&-0.000139&-0.000293&-0.000129&$2.432\times10^{-3}$&$2.432\times10^{-3}$&$2.21\times10^{11}$
\\\hline 37&$J/\psi+D_{s0}^{*-}$
&0.000458&0.00108&0.000427&$2.299\times10^{-4}$&$2.297\times10^{-4}$&$2.15\times10^{11}$
\\\hline 38&$J/\psi+D_s^{*-}$
&-0.000884&-0.00196&-0.000824&$6.752\times10^{-3}$&$6.764\times10^{-3}$&$1.96\times10^{9}$
\\\hline 39&$J/\psi+D_{s1}^-(2460)$
&-0.000875&-0.00192&-0.000817&$5.339\times10^{-3}$&$5.348\times10^{-3}$&$2.53\times10^{9}$
\\\hline 40&$J/\psi+D_{s1}^-(2536)$
&-0.000639&-0.00151&-0.000596&$1.123\times10^{-3}$&$1.124\times10^{-3}$&$2.26\times10^{10}$
\\\hline
\end{tabular} \label{tab5}}
\end{table}

\begin{table}[ph]
\caption{The CP asymmetries and branching ratios of $B_c$
non-leptonic decays [\uppercase\expandafter{\romannumeral2}]. $D_1$
and $D_2$ are defined as $\mathcal{A}_{CP}\equiv
D_1\frac{\sin\gamma}{1+D_2\cos\gamma}$, where $\gamma$ is the weak
phase (see Eq. (\ref{eq1})); $\epsilon_f N$ are the numbers of
$B_c^{\pm}$ events needed for testing CP violation at three standard
deviation (3$\sigma$) level, which are decided by $\epsilon_f
N\sim\frac{9}{Br \mathcal{A}^2_{cp}}$, where $\epsilon_f$ is the
detecting efficiency of the final state. The following values are
taken in calculation: $a_1=1.14$, $a_2=-0.2$ and weak phase
$\gamma=68.8^{\circ}$. Color-favored tree and penguin diagrams
contribute in the channels (41-52); color-suppressed tree and
penguin diagrams contribute in the channels (59-64); for the other
channels in this table, only penguin diagram contributes.}
{\begin{tabular}{|c|c|c|c|c|c|c|c|} \hline No.&Final
States&$D_1$&$D_2$&$\mathcal{A}_{cp}$&$B_r(B_c^+\rightarrow
\bar{f})$&$B_r(B_c^-\rightarrow f)$&$\epsilon_f N$
\\\hline 41&$\bar{D}^0+K^-$
&0.133&-0.509&0.152&$1.83\times10^{-7}$&$1.34\times10^{-7}$&$2.44\times10^{9}$
\\\hline 42&$\bar{D}^0+K^{*-}$
&0.201&-0.735&0.255&$1.70\times10^{-7}$&$1.01\times10^{-7}$&$1.02\times10^{9}$
\\\hline 43&$\bar{D}^{*0}+K^-$
&0.235&-0.970&0.338&$5.41\times10^{-8}$&$2.68\times10^{-8}$&$1.95\times10^{9}$
\\\hline 44&$\bar{D}^{*0}+K^{*-}$
&0.201&-0.735&0.255&$3.91\times10^{-7}$&$2.32\times10^{-7}$&$4.44\times10^{8}$
\\\hline 45&$\bar{D}_0^{*0}+K^-$
&0.235&-0.970&0.338&$2.31\times10^{-8}$&$1.14\times10^{-8}$&$4.57\times10^{9}$
\\\hline 46&$\bar{D}_0^{*0}+K^{*-}$
&0.201&-0.735&0.255&$1.56\times10^{-7}$&$9.23\times10^{-8}$&$1.11\times10^{9}$
\\\hline\hline 47&$\bar{D}^0+\pi^-$
&-0.104&0.338&-0.0861&$1.93\times10^{-7}$&$2.30\times10^{-7}$&$5.75\times10^{9}$
\\\hline 48&$\bar{D}^0+\rho^-$
&-0.0640&0.221&-0.0552&$5.41\times10^{-7}$&$6.04\times10^{-7}$&$5.15\times10^{9}$
\\\hline 49&$\bar{D}^{*0}+\pi^-$
&-0.0285&0.0992&-0.0256&$4.25\times10^{-7}$&$4.47\times10^{-7}$&$3.15\times10^{10}$
\\\hline 50&$\bar{D}^{*0}+\rho^-$
&-0.0640&0.221&-0.0552&$1.22\times10^{-6}$&$1.36\times10^{-6}$&$2.29\times10^{9}$
\\\hline 51&$\bar{D}_0^{*0}+\pi^-$
&-0.0285&0.0992&-0.0256&$1.77\times10^{-7}$&$1.86\times10^{-7}$&$7.56\times10^{10}$
\\\hline 52&$\bar{D}_0^{*0}+\rho^-$
&-0.0640&0.221&-0.0552&$4.93\times10^{-7}$&$5.51\times10^{-7}$&$5.65\times10^{9}$
\\\hline\hline 53&$D^-+\bar{K}^0$
&0.00593&0.0375&0.00545&$1.91\times10^{-7}$&$1.89\times10^{-7}$&$1.60\times10^{12}$
\\\hline 54&$D^-+\bar{K}^{*0}$
&0.00576&0.0363&0.00530&$1.60\times10^{-7}$&$1.58\times10^{-7}$&$2.01\times10^{12}$
\\\hline 55&$D^{*-}+\bar{K}^0$
&0.00593&0.0375&0.00545&$3.20\times10^{-8}$&$3.16\times10^{-8}$&$9.52\times10^{12}$
\\\hline 56&$D^{*-}+\bar{K}^{*0}$
&0.00576&0.0363&0.00530&$3.71\times10^{-7}$&$3.68\times10^{-7}$&$8.68\times10^{11}$
\\\hline 57&$D_0^{*-}+\bar{K}^0$
&0.00593&0.0375&0.00545&$1.36\times10^{-8}$&$1.35\times10^{-8}$&$2.24\times10^{13}$
\\\hline 58&$D_0^{*-}+\bar{K}^{*0}$
&0.00576&0.0363&0.00530&$2.72\times10^{-7}$&$2.70\times10^{-7}$&$1.18\times10^{12}$
\\\hline\hline 59&$D^-+\pi^0$
&-0.419&0.884&-0.296&$3.88\times10^{-9}$&$7.14\times10^{-9}$&$1.87\times10^{10}$
\\\hline 60&$D^-+\rho^0$
&-0.359&0.909&-0.252&$9.44\times10^{-9}$&$1.58\times10^{-8}$&$1.13\times10^{10}$
\\\hline 61&$D^{*-}+\pi^0$
&-0.248&0.713&-0.184&$6.42\times10^{-9}$&$9.31\times10^{-9}$&$3.39\times10^{10}$
\\\hline 62&$D^{*-}+\rho^0$
&-0.359&0.909&-0.252&$2.11\times10^{-8}$&$3.52\times10^{-8}$&$5.05\times10^{9}$
\\\hline 63&$D_0^{*-}+\pi^0$
&-0.248&0.713&-0.184&$2.70\times10^{-9}$&$3.92\times10^{-9}$&$8.07\times10^{10}$
\\\hline 64&$D_0^{*-}+\rho^0$
&-0.359&0.909&-0.252&$8.43\times10^{-9}$&$1.41\times10^{-8}$&$1.26\times10^{10}$
\\\hline\hline 65&$D_s^{-}+K^0$
&-0.0593&-0.650&-0.0723&$2.19\times10^{-8}$&$2.54\times10^{-8}$&$7.27\times10^{10}$
\\\hline 66&$D_s^{-}+K^{*0}$
&-0.0747&-0.633&-0.0904&$1.90\times10^{-8}$&$2.28\times10^{-8}$&$5.28\times10^{10}$
\\\hline 67&$D_{s0}^{*-}+K^0$
&-0.0593&-0.650&-0.0723&$1.03\times10^{-9}$&$1.19\times10^{-9}$&$1.55\times10^{12}$
\\\hline 68&$D_{s0}^{*-}+K^{*0}$
&-0.0747&-0.633&-0.0904&$1.08\times10^{-8}$&$1.30\times10^{-8}$&$9.25\times10^{10}$
\\\hline 69&$D_s^{*-}+K^0$
&-0.0593&-0.650&-0.0723&$3.22\times10^{-9}$&$3.73\times10^{-9}$&$4.95\times10^{11}$
\\\hline 70&$D_s^{*-}+K^{*0}$
&-0.0747&-0.633&-0.0904&$3.75\times10^{-8}$&$4.50\times10^{-8}$&$2.67\times10^{10}$
\\\hline 71&$D_{s1}^{-}(2460)+K^0$
&-0.0593&-0.650&-0.0723&$3.75\times10^{-8}$&$4.34\times10^{-8}$&$4.25\times10^{10}$
\\\hline 72&$D_{s1}^{-}(2460)+K^{*0}$
&-0.0747&-0.633&-0.0904&$3.68\times10^{-8}$&$4.41\times10^{-8}$&$2.72\times10^{10}$
\\\hline 73&$D_{s1}^{-}(2536)+K^0$
&-0.0593&-0.650&-0.0723&$3.61\times10^{-8}$&$4.17\times10^{-8}$&$4.43\times10^{10}$
\\\hline 74&$D_{s1}^{-}(2536)+K^{*0}$
&-0.0747&-0.633&-0.0904&$3.15\times10^{-8}$&$3.77\times10^{-8}$&$3.18\times10^{10}$
\\\hline\hline 75&$D_s^{-}+\phi^0$
&0.00422&0.0414&0.00387&$5.19\times10^{-7}$&$5.15\times10^{-7}$&$1.16\times10^{12}$
\\\hline 76&$D_{s0}^{*-}+\phi^0$
&0.00422&0.0414&0.00387&$3.02\times10^{-7}$&$3.00\times10^{-7}$&$1.99\times10^{12}$
\\\hline 77&$D_s^{*-}+\phi^0$
&0.00422&0.0414&0.00387&$1.05\times10^{-6}$&$1.04\times10^{-6}$&$5.73\times10^{11}$
\\\hline 78&$D_{s1}^{-}(2460)+\phi^0$
&0.00422&0.0414&0.00387&$1.08\times10^{-6}$&$1.07\times10^{-6}$&$5.56\times10^{11}$
\\\hline 79&$D_{s1}^{-}(2536)+\phi^0$
&0.00422&0.0414&0.00387&$8.65\times10^{-7}$&$8.59\times10^{-7}$&$6.95\times10^{11}$
\\\hline
\end{tabular} \label{tab6}}
\end{table}

The channels of $B_c$ decaying into two heavy mesons $i.e.$ a charmonium
and a $D$ or $D_s$ meson are listed in Table \ref{tab5}. These
decays are dominated by the tree diagrams, and only for CP
violation, the penguin diagram effects arise through the
interference with the tree diagram. The decay ratios of the
processes with a $D$ meson (channels 1-15) in the final state are
generally smaller than those with a $D_s$ meson (channels 16-40),
since in the former processes the tree diagrams have CKM factor
$V_{cb}V^*_{cd}$ which is of order $\lambda^3$ while in the
processes (16-40) the tree diagrams have CKM factor
$V_{cb}V^*_{cs}\sim \lambda^2$, where $\lambda=0.2253$~\cite{PDG} is
a Wolfenstein parameter. The CP asymmetries in the
channels (1-15) are generally larger than those in (16-40). In order
to test the CP violating effects, we need both branching ratio and
CP asymmetry to be sufficiently large. It is shown in the Table
\ref{tab5} that the most favorite channels are $B_c^-\rightarrow \eta_c
+D^-/D^{*-}_0$.

In Table \ref{tab6}, we show that color-favored tree and penguin diagrams
contribute in the channels (41-52); color-suppressed tree and
penguin diagrams contribute in the channels (59-64); for the other
channels in this table, only penguin diagram contributes. As
discussed before, in the amplitudes of the processes (41-46) and
(59-64), the tree contribution and the penguin contribution are in
the same order, so only (47-52) are tree dominated processes. The
branching ratios in Table \ref{tab6} are generally smaller than those in
Table \ref{tab5}: It is because (i) the tree amplitudes (if exist) in the
channels of Table \ref{tab6} are Cabibbo suppressed due to the small
magnitude of $V_{ub}$, and  (ii)  most of the form factors
for the decays listed in Table \ref{tab6} are smaller than those in
Table \ref{tab5}.

The processes (41-46) and (59-64) have significantly large CP
asymmetries. The CP violating effects are mainly coming from the
interference between the tree diagram and the penguin diagram. Since
the branching ratios of these decays are small, the numbers
of $B_c$ for testing CP effects are around $10^9\sim10^{10}$, which
may be too large for LHC experiments.

\begin{figure}[hb]
\centering
\includegraphics[width = 0.8\textwidth]{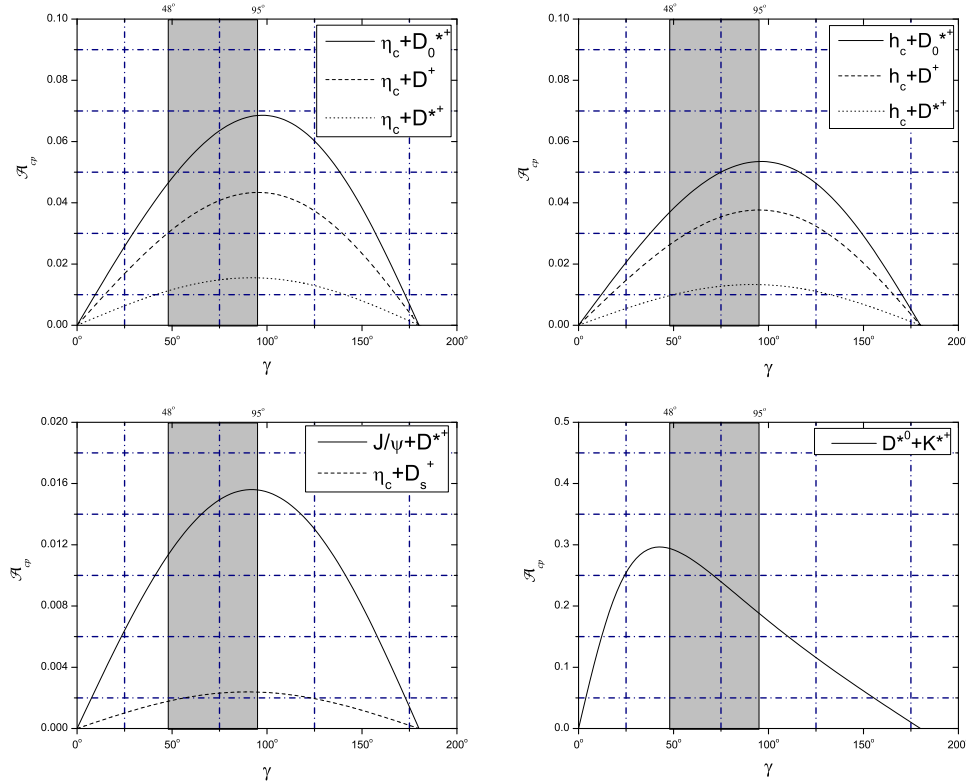}
\caption{Dependence of the CP asymmetries $\mathcal{A}_{CP}$ upon
the weak phase $\gamma$ in a few interesting processes. The shaded region
is the constrained range of $\gamma$ measured in tree level $B$
decay, which is about $48^\circ\sim95^\circ$.} \label{fig2}
\end{figure}

According to Table \ref{tab5} and \ref{tab6}, the $\epsilon_f N$ of
the decays $B_c^-\rightarrow \eta_c +D^-/D^{*-}_0$ are of order
$\sim10^7$ and the values for the processes $B_c^-$ decaying into $
\eta_c +D^{*-}/D_s^-, \ h_c+D^-/D^{*-}/D^{*-}_0,\ J/\Psi+D^{*-}$ and
$\bar{D}^{*0}+K^{*-}$ are of order $\sim10^8$. Since the LHC-b is
expected to produce around $5\times10^{10}$ $B_c$ events per year,
they are hopeful channels to be tested for the CP violation. In
Fig.~\ref{fig2}, we draw the CP asymmetry $vs$ the weak phase of
these channels to see their dependence on $\gamma$ using
Eq.~(\ref{eq1}). The shaded region is the constrained range of
$\gamma$ measured in tree level $B$ decay, which is
$48^\circ\sim95^\circ$~\cite{PDG}. From the figure we can see  the
CP asymmetries of these processes may suffer about $\sim30\%$
uncertainty. Fortunately the Global fit of the Wolfenstein
parameters provided a more rigorous constraint for the weak phase
$\gamma$.

\begin{table} \small\caption{The branching ratios and CP asymmetries
of  a few interesting channels at different $k^2$, where $k$ is the momentum
of gluon in the QCD loop integrals (see Eqs.~(\ref{eq5}-\ref{eq2})).}
\begin{tabular}{|c|c|c|c|c|c|c|c|c|}
\hline &\multicolumn{4}{|c|}{$B_r(B_c^+\rightarrow
\bar{f})$}&\multicolumn{4}{|c|}{$\mathcal{A}_{CP}$}\\
\hline
$k^2$&0.35$m_b^2$&0.50$m_b^2$&0.65$m_b^2$&0.80$m_b^2$&0.35$m_b^2$&0.50$m_b^2$&0.65$m_b^2$&0.80$m_b^2$\\
\hline
$\eta_c+D^+$&$1.09\times10^{-4}$&$1.13\times10^{-4}$&$1.15\times10^{-4}$&$1.16\times10^{-4}$&0.0383&0.0391&0.0389&0.0387\\
\hline
$\eta_c+D_0^{*+}$&$3.89\times10^{-5}$&$4.10\times10^{-5}$&$4.20\times10^{-5}$&$4.27\times10^{-5}$&0.0593&0.0609&0.0605&0.0599\\
\hline
$\eta_c+D^{*+}$&$1.16\times10^{-4}$&$1.18\times10^{-4}$&$1.18\times10^{-4}$&$1.19\times10^{-4}$&0.0142&0.0143&0.0143&0.0143\\
\hline
$h_c+D^+$&$6.41\times10^{-5}$&$6.59\times10^{-5}$&$6.69\times10^{-5}$&$6.75\times10^{-5}$&0.0332&0.0338&0.0337&0.0335\\
\hline
$h_c+D_0^{*+}$&$1.20\times10^{-5}$&$1.25\times10^{-5}$&$1.28\times10^{-5}$&$1.30\times10^{-5}$&0.0469&0.0480&0.0477&0.0473\\
\hline
$h_c+D^{*+}$&$7.08\times10^{-5}$&$7.16\times10^{-5}$&$7.20\times10^{-5}$&$7.23\times10^{-5}$&0.0122&0.0123&0.0123&0.0123\\
\hline
$J/\psi+D^{*+}$&$3.17\times10^{-4}$&$3.21\times10^{-4}$&$3.23\times10^{-4}$&$3.24\times10^{-4}$&0.0143&0.0144&0.0144&0.0143\\
\hline
$\eta_c+D_s^+$&$2.69\times10^{-3}$&$2.79\times10^{-3}$&$2.84\times10^{-3}$&$2.88\times10^{-3}$&-0.00223&-0.00226&-0.00225&-0.00223\\
\hline
$D^{*0}+K^{*+}$&$4.02\times10^{-7}$&$3.85\times10^{-7}$&$3.66\times10^{-7}$&$3.51\times10^{-7}$&0.142&0.279&0.342&0.383\\
\hline
\end{tabular}\label{tab7}
\end{table}

In calculating the QCD loop integrals we take the gluon momentum $k$
by using Eq.~(\ref{eq2}). For the processes in Table \ref{tab5}, $k^2$
is estimated to be around $(0.6\sim0.7)m_b^2$, and for the processes in Table
\ref{tab6}, $k^2$ to be around $(0.4\sim0.5)m_b^2$. Eq.~(\ref{eq2})
is based on a simple kinematic picture, but it is too simple to
reflect the final state hadronization dynamics. So we investigate
how much it will affect the CP asymmetry and decay rate if the value
of $k^2$ varies. The branching ratios and CP asymmetries of most
interested channels at different $k^2$ are shown in Table
\ref{tab7}. According to the table, the branching ratios of all the
channels are barely affected by $k^2$. For CP asymmetries, only the
CP asymmetry of $B_c^-\rightarrow\bar{D}^{*0}+K^{*-}$ process is
sensitive to the value of $k^2$. Actually we find that the
branching ratios of all the channels calculated are barely affected by
the value of $k^2$, and so are the CP asymmetries of processes
(1-40), but the CP asymmetries of processes (41-79) are sensitive to
$k^2$.

In summary, we calculated the decay rates and CP asymmetries of non-leptonic two-body decay of $B_c$ meson.
Based on our calculation we have found
that the best decay channels to observe CP violation  at LHC
are $B_c^-\rightarrow \eta_c +D^-/D^{*-}_0$. Decays to $ \eta_c
+D^{*-}/D_s^-, \ h_c+D^-/D^{*-}/D^{*-}_0,\ J/\Psi+D^{*-}$ are also
hopeful channels. In this work the soft strong phases are not
calculated, which may cause uncertainty of the results and need
further study.

\section*{Acknowledgments}

The work of C.S.K. was  supported in part by the Basic Science Research
Program through the NRF of Korea funded by MOEST (2009-0088395) and
in part by KOSEF through the Joint Research Program
(F01-2009-000-10031-0).
The work of G.W. was supported in part by the National Natural Science
Foundation of China (NSFC) under Grant No. 10875032 and supported
in part by Projects of International Cooperation and Exchanges
NSFC under Grant No. 10911140267.

\end{document}